\pdfoutput=1

\documentclass[11pt]{article}

\usepackage[]{emnlp2021}

\usepackage{times}
\usepackage{latexsym}

\usepackage[T1]{fontenc}

\usepackage[utf8]{inputenc}
\usepackage{natbib}

\usepackage{colortbl}
\usepackage{subfigure}
\usepackage{
  float,
  epsfig,
  wrapfig,
  graphics
  }
\usepackage{graphicx}
\usepackage{comment}
\usepackage{booktabs}
\newcommand{\ourdataset}{\texttt{CA4P-483}}
\newcommand{\ourtask}{$\texttt{CSP}^3$}
\usepackage{hhline}
\usepackage{
  tikz,
  pgfplots,
  pgfplotstable
}
\usepackage{enumitem}
\usepackage{xcolor}
\usepackage{microtype}

\usepackage{amsmath,amsfonts,amsthm}
\usepackage[hang,flushmargin]{footmisc}

\usepackage{tcolorbox}

\usepackage{CJKutf8}

\title{A Fine-grained Chinese Software Privacy Policy Dataset for Sequence Labeling and Regulation Compliant Identification}

\author{Kaifa Zhao$^1$, Le Yu$^1$, Shiyao Zhou$^1$, Jing Li$^1$, Xiapu Luo$^{1}$, \\ \textbf{Yat Fei Aemon Chiu}$^2$, \textbf{Yutong Liu}$^2$ \\
\small{$^1$Department of Computing, The Hong Kong Polytechnic University, HKSAR, China} \\
\small{$^2$Department of Electronic and Information Engineering, The Hong Kong Polytechnic University, HKSAR, China} \\
\small{$^1$\texttt{kaifa.zhao@connect.polyu.hk, lele08.yu@polyu.edu.hk, shiyao.zhou@connect.polyu.hk}}\\
\small{$^1$\texttt{\{jing-amelia.li, daniel.xiapu.luo\}@polyu.edu.hk}} \\ \small{$^2$\texttt{\{yat-fei-dylan.zhao,yitang.liu\}@connect.polyu.hk}}
}

\begin{document}
\maketitle
\begin{abstract}

Privacy protection raises great attention on both legal levels and user awareness.
To protect user privacy, countries enact laws and regulations requiring software privacy policies to regulate their behavior.
However, privacy policies are written in natural languages with many legal terms and software jargon that prevent users from understanding and even reading them.
%
It is desirable to use NLP techniques to analyze privacy policies for helping users understand them.
%
Furthermore, existing datasets ignore law requirements and are limited to English.
In this paper, we construct the first Chinese privacy policy dataset, namely {\ourdataset}, to facilitate the sequence labeling tasks and regulation compliance identification  between privacy policies and software.
Our dataset includes 483 Chinese Android application privacy policies, over 11K sentences, and 52K fine-grained annotations. 
We evaluate families of robust and representative baseline models on our dataset.
Based on baseline performance, we provide findings and potential research directions on our dataset. 
Finally, we investigate the potential applications of {\ourdataset}\footnote{Our dataset and code are publicly available in \url{https://github.com/zacharykzhao/CA4P-483}} combing regulation requirements and program analysis.




\end{abstract}
\section{Introduction}






A privacy policy is a legal document written in natural language that discloses how and why a controller collects, shares, uses, and stores user data~\cite{gdpr,piss,zhSDKLaw}.
%
Privacy policies help users understand whether their privacy will be abused and decide whether to use the product.
However, privacy policies are tedious, making it hard for users to read and understand them~\cite{staff2011protecting}.
Natural language processing techniques achieve big success on understanding document semantics~\cite{yang2021time,wen2021automatically,ding2020hashtags}.
Thus, it is necessary to apply natural language processing to analyze the privacy policies~\cite{yu2016can,ppcheck,le2015ppg,fan2020empirical} and help users be aware of apps' privacy access behavior~\cite{zhou2021finding}.

\textbf{C}hinese \textbf{s}oftware \textbf{p}rivacy \textbf{p}olicy \textbf{p}rocessing ({\ourtask}) task is a sequence labeling problem that 
recognizes privacy-related components in the sentences.
%
{\ourtask} has two main unique features.
First, privacy policies contain an amount of information inside~\cite{ppchecker}, such as how the app stores user data, and how to contact app developer.
In our dataset, we concentrate on data access-related sentences as the sentences directly related to user privacy.
Second, privacy policies are written in a legally binding professional language and contain software jargon.
Thus, it requires strong background~\cite{hao22ccs,zhou2022uncovering} to understand the statements inside.
Both characteristics prevent users from understanding the privacy policies.
%
A well-annotated dataset can facilitate building automatic privacy policy analysis tool and further help users protect their privacy.

%

Although privacy policy datasets have been proposed recently~\cite{wilson2016creation, zimmeck2019maps}, 
labels in existing datasets are coarse-grained (i.e., sentence-level  annotations~\cite{wilson2016creation}) and limited to few privacy practices~\cite{zimmeck2019maps}.
%
Besides, existing datasets only include English privacy policies, which limits the application of these datasets in regions with other languages.
%
We construct a  fine-grained Chinese dataset for software privacy policy analysis.

In this work, we focus on Android application privacy policies as Android possesses the largest share of mobile operating systems ~\cite{osmarket} and a large number of Android privacy data leaks have been revealed~\cite{shrivastava2021android, sivan2019analysis}.
Unlike previous work~\cite{wilson2016creation,zimmeck2019maps}, we deal with the problem using sequence labeling methods, 
and pay special attention to Chinese privacy policies.
The motivations come from the following four aspects:

First, worldwide regulation departments enact laws~\cite{zhSDKLaw, piss, gdpr, ccpa, clprc} to regulate the software's behaviors and protect users' privacy.
The laws require the software to clarify how and why they need to access user data.
Analyzing privacy policies can help users understand how app process their data and identify whether apps comply with laws.
%
%
Second, for sequence labeling tasks,  {\ourtask} aims to identify how and why the software collects, shares, and manages users' data according to regulations.
{\ourtask} can be abstracted as identifying components in the privacy policy documents, such as data type and the purpose of using user data.
NLP techniques can help automatically analyze privacy policies.
Third,  existing privacy policy analysis research is limited to English and totally omits other languages.
With over 98.38 billion app downloads~\cite{statista} and privacy-related regulations 
enacted in China, it is necessary and urgent to research {\ourtask}. 
%
Last but not least, recent research in other communities, such as software engineering~\cite{yu2016can,icse2022preksha} and cyber security~\cite{ppcheck,andow2019policylint}, demonstrates requirements for analyzing privacy policies to help the analyst identify whether the apps' behavior is consistent with privacy policies.

In this work, we make the following efforts to advance {\ourtask}:

First, we construct a novel large-scale human-annotated \textbf{C}hinese \textbf{A}ndroid a\textbf{pp}lication \textbf{p}rivacy \textbf{p}olicy dataset, namely {\ourdataset}.
%
%
Specifically, we manually visit the software markets, such as Google Play~\cite{googleplay} and AppGallery~\cite{appgallery}, check the provided privacy policy website, and download the Chinese version if available.
We finally collect 483 documents.
To determine the labels in the privacy policy analysis scenario, we read through Chinese privacy-related regulations and summarize seven components (\S\ref{sec:annotation_format}).
We annotate all occurrences of components in 11,565 sentences from 483 documents.
Unlike paragraph-level annotations in existing privacy policy datasets~\cite{wilson2016creation}, {\ourdataset} annotates character-level corpus.

Second, based on {\ourdataset}, we summarize families of representative baselines for Chinese sequence labeling.
In detail, we first evaluate the performance of several classic sequence labeling models on our dataset, including Conditional Random Forest (CRF)~\cite{kudo2005crf},
Hidden Markov Model (HMM)~\cite{morwal2012named},
BiLSTM~\cite{graves2005framewise},
BiLSTM-CRF~\cite{lample2016neural}, 
and BERT-BiLSTM-CRF~\cite{devlin2018bert}.
%
Recent work shows lattice knowledge improves the performance of Chinese sequence labeling tasks.
We involve lexicon-based models, such as Lattice-LSTM~\cite{zhang2018chinese}.
%
%

Third, we investigate potential applications of {\ourdataset}.
%
Combining law knowledge, we first identify whether the privacy policy violates regulation requirements based on {\ourdataset}.
We also identify whether the app behaves consistently with privacy policy statements combing software analysis~\cite{zhao2021structural,zhou2020demystifying}.

The contributions of this work are three-fold: 
\begin{itemize}[leftmargin=*]
    \vspace{-0.5em}

\item  To the best of our knowledge, we construct the first Chinese privacy policy dataset, namely CA4P-483, integrating abundant fine-grained annotations.
%

    \vspace{-0.5em}
\item We experimentally evaluate and analyze the results of different families of sequence labeling baseline models on our dataset.
%
We also summarize difficulties in our dataset, and provide findings and further research topics on our dataset.

    \vspace{-0.5em}
\item We investigate potential applications of {\ourdataset} to regulate privacy policies with law knowledge and program analysis technologies.
    
    
\end{itemize}

\section{Dataset Construction}
\label{sec:data_construction}

\begin{figure*}[t]
    \centering
    \subfigure[Demo 1.]{
        \centering
        \includegraphics[width=0.95\textwidth]{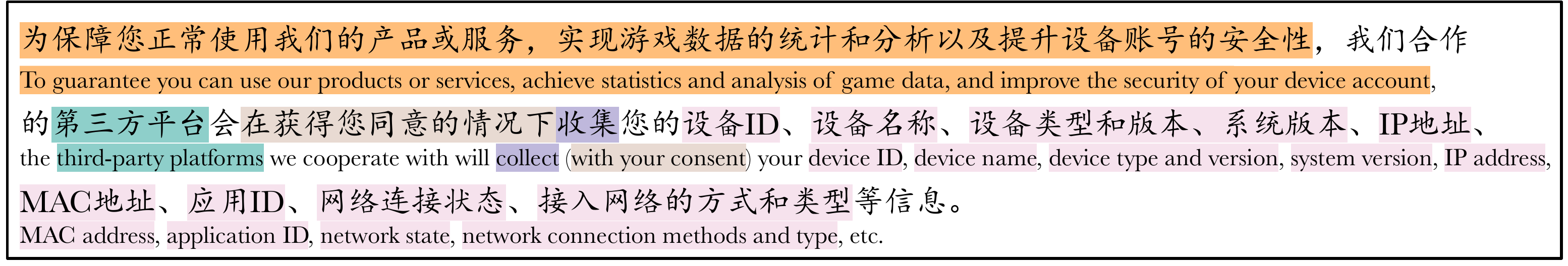}
        \label{fig:ann_demo1}
     }
    \subfigure[Demo 2.]{
        \centering
        \includegraphics[width=0.95\textwidth]{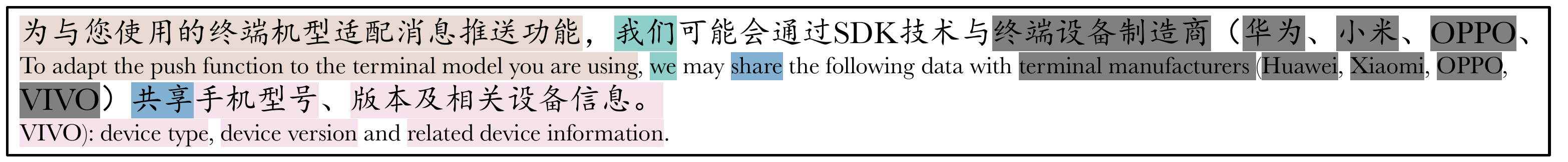}
        \label{fig:ann_demo2}
    }    
    \subfigure[Annotation legend.]{
        \centering
        \includegraphics[width=0.95\textwidth]{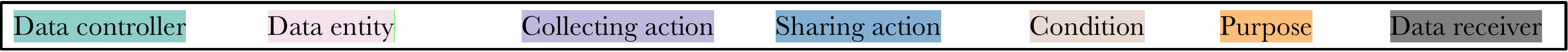}
        \label{fig:ann_legend}
    }
   \caption{Annotation demos from {\ourdataset}. We translate the statements into English for illustration.}
   \label{fig:ann_demo}
   
\end{figure*}

\subsection{Dataset collection}
\begin{CJK}{UTF8}{gbsn}


We manually collect the Chinese privacy policies from Android application markets.
%
According to application market requirements~\cite{appgallery_requirements,googleplay_requirements}, developers \textbf{must} provide privacy policies to claim their user data access behavior and to ensure apps will not violate laws or regulations.
Since privacy policies are publicly available for users to understand the apps' access of personal data, three authors of this paper manually access the most popular apps in markets and visit their privacy policy websites provided at the moment (January 2021).
We use \textit{html2text}~\cite{html2text} to extract context.
Finally, we use \textit{tagtog}~\cite{tagtog} for document annotation.




Next, we annotate {\ourdataset} based on the law requirements.
Specifically, we analyze Chinese privacy-related laws and regulations \cite{zhSDKLaw, piss, MDICUPIA, bscp}, and find requirements for apps' privacy process behavior.  
For example, GB/T41391-2022 Article 4.n) claims that "\textit{developers should expressly state the \underline{purpose} of applying or \underline{collecting} information to the \underline{subject} of \underline{personal information}}."
Finally, we summarize seven types of labels related to requirements for apps' access to user data.

\end{CJK}

\subsection{Fine-grained annotations}
\label{sec:annotation_format}

For each privacy policy, we concentrate on the sentences that describe the data process behavior.
After locating the sentences, we annotate seven components, i.e., the controller, data entity, collection, sharing, condition, purpose, and receiver.

\paragraph{Data controller.}
According to regulation requirements, the data controller is the party that determines the purpose and means of personal data processing. 
A data controller could be the app (first party) or the third party.
As is shown in Fig.\ref{fig:ann_demo}, data controllers are "\textit{third-party platforms}" in Fig.\ref{fig:ann_demo1} while that is "\textit{we}" in Fig.\ref{fig:ann_demo2}.
Thus, we annotate data controllers according to sentence semantics, i.e., who is responsible for processing the data.


\paragraph{Data entity.}
Data entities are any information that can identify or reflect the activities of a natural person~\cite{piss}.
Recent research~\cite{cai2012practicality,shokri2017membership} demonstrates the probability of combining various information to infer and even locate a specific person.
Thus, we annotate all data nouns or noun phrases that are requested in privacy policies, including sensitive information, such as device id, and normal information, such as device type.

\paragraph{Collection.}
\begin{CJK}{UTF8}{gbsn}
Collection actions are verbs that describe how controllers access data, such as gather (收集), obtain (获取).

\end{CJK}

\paragraph{Sharing.}
Sharing actions are verbs that indicate whether the data controller will distribute data to others. 
Although both Sharing and Collection describe how the party access user data, we difference them according to the requirements of laws on the action, such as Article 5 and 9.2 in  ~\cite{piss}.

\paragraph{Condition.}
Condition describes the situation where the data controller will access personal data.
%
Laws require data controllers to inform users under what conditions their data will be processed.
For example, bank apps may require the users' identification information when activating bank account.
%
%

\paragraph{Purpose.}
Purpose should claim why the data controller processes user data.
Laws enact specific requirements for user data access.
For example, PISS Article 4.d) requires 
controllers to clearly state purpose of processing data.
Purpose can also help the users understand why the app collects their data and further determine whether to give the consent as is shown in Fig.\ref{fig:ann_demo1}.

\paragraph{Data receiver.}
Data receiver describes the parties that receive user data.
Laws not only ask apps to clarify who will get shared data ~\cite{piss} but also restrict the data receivers' behavior~\cite{zhSDKLaw}, such as why processing user data.
%

\vspace{-0.5em}
\subsection{Human annotation process}
\vspace{-0.5em}

Our privacy policy annotation consists of two phases: coarse-grained annotation and fine-grained annotation.
Coarse-grained annotation labels privacy policies at paragraph level following previous work~\cite{wilson2016creation}.
Fine-grained annotation labels our defined components at the word level based on coarse-grained annotation.


For the first phase, three authors of this paper, who have researched privacy policies and software engineering for over eight and three years, label ten privacy policies for reference and record a video instruction to guide annotators.
Then, we hire thirty undergraduates in our university to annotate the dataset. 
The three instructors train each annotator for at least four hours to be familiar with the dataset and requirements.
Students are asked to annotate 1000 Android apps' privacy policies in Chinese and each privacy policy should be analyzed for at least 30 minutes to ensure quality.
Each privacy policy is allocated to at least four annotators.
Finally, three instructors inspect each annotation.

For the second phase, we select two undergraduates, who coarse-grained annotate the documents with high precision, to conduct the fine-grained annotation.
Specifically, we select 483 documents that are well coarse-grained annotated after inspection.
Instructors first annotate ten documents to lead undergraduates to annotate.
%
%
The annotators also keep discussing with instructors once the role of components in sentences are unclear.
%
Each annotator is required to label each privacy policy for at least 30 minutes to guarantee the dataset quality.

Finally, the instructors analyze the annotations and use Fleiss' Kappa metrics~\cite{cohen1960coefficient, wilson2016creation} to evaluate the agreements. 
Table~\ref{tab:data_statistic} shows the average Kappa value (77.20\%) satisfies the substantial agreement, i.e., Kaapa value lies in 0.61-0.80, 
and four components achieve almost perfect agreement (0.81-1.00).
The \textit{Condition}, which only gets moderate agreement, is caused by the overlap between labels (details in Appendix~\ref{app:demo}).
%
%



\begin{table}[]
\centering
\resizebox{0.99\linewidth}{!}{%
\vspace{1em}
\begin{tabular}{@{}rrrrrrr@{}}
\toprule
\multicolumn{4}{c}{\# doc}                & \multicolumn{3}{c}{483}    \\ \midrule
\multicolumn{4}{c}{\# sentences}          & \multicolumn{3}{c}{11,565} \\ \midrule
\multicolumn{4}{c}{\# sentences with ann} & \multicolumn{3}{c}{3,385}  \\ \midrule
\multicolumn{4}{c}{Avg sentences len}  & \multicolumn{3}{c}{79.06}  \\ 
\bottomrule
Type           & Num    & \multicolumn{1}{c}{Train}  & \multicolumn{1}{c}{Dev}   & \multicolumn{1}{c}{Test}  & \multicolumn{1}{c}{Avg len} & \multicolumn{1}{c}{Kappa} \\ 
\hline \hline
Data            & 21,241     & 18,925     & 2,521   & 2,331   & 4.68 & 85.39\%  \\
Collect         & 5,134      & 4,133      & 576     & 528     & 2.03 & 73.78\%  \\
Share           & 4,976      & 3,989      & 533     & 505     & 2.10 & 84.87\% \\
Controller      & 8,424      & 6,085      & 815     & 782     & 2.49 &  82.22\% \\
Condition       & 4,917      & 5,477      & 716     & 713     & 14.41 &  50.07\%    \\
Receiver        & 3,202      & 2,776      & 360     & 350     & 4.29  & 89.88\% \\
Purpose         & 4,683      & 6,442      & 860     & 867     & 19.24 & 74.18\%     \\ \midrule
Total & 52,577 & \multicolumn{1}{c}{47,827} & \multicolumn{1}{c}{6,381} & \multicolumn{1}{c}{6,076} &          \\ \bottomrule
\end{tabular}%
}
\caption{The statistics of {\ourdataset}. Here, "Avg" denotes \textit{average}, "ann" denotes \textit{annotation}, "len" denotes \textit{length}, "\#" denotes \textit{the number of}.}
\label{tab:data_statistic}
\end{table}

\subsection{Dataset statistics and comparison}



We conduct statistical analysis and show the results in Table~\ref{tab:data_statistic}.
{\ourdataset} is split into
training, development, and test set.
Table~\ref{tab:data_statistic} also gives details of the number of different labels in each set.
Table~\ref{tab:data_statistic} shows that the average length of condition and purpose is much longer than other corpora as the two types are generally in the form of clauses.

We compare {\ourdataset} with related datasets in Table~\ref{tab:data_set_comparison}.
We first compare our corpus with Chinese sequence labeling datasets, such as MSRA~\cite{zhang2006word}, 
OntoNotes~\cite{weischedel2011ontonotes}, 
Weibo~\cite{peng2016improving}, 
PeopleDiary~\cite{peopodaily},
Resume~\cite{zhang2018chinese},
CLUENER2020~\cite{xu2020cluener2020}, and
CNERTA~\cite{sui2021large}.
We also involve widely used English sequence labeling datasets, namely Twitter-2015~\cite{zhang2018adaptive} and Twitter-2017~\cite{lu2018visual}.
We also consider privacy policy datasets, namely Online Privacy Policies (OPP-115) ~\cite{wilson2016creation} and Android app privacy policies (APP-350)~\cite{zimmeck2019maps}.

\begin{table*}[t]
\centering
\resizebox{0.75\textwidth}{!}{%
\begin{tabular}{@{}crrrrcc@{}}
\toprule
Dataset       & \multicolumn{1}{c}{\# Train}        & \multicolumn{1}{c}{\# Dev}     & \multicolumn{1}{c}{\# Test}     & \multicolumn{1}{c}{Size}      & Language & \# Class \\ 
\hline\hline
MSRA          & 41,728          & 4,636      & 4,365       & 50K       & Chinese  & 3 \\
PeopleDairy   & 20,864          & 2,318      & 4,636       & 23k       & Chinese  & 3 \\
Weibo         & 1,350           & 270        & 270         & 2k        & Chinese  & 4 \\
Resume        & 3,821           & 463        & 477         & 2k        & Chinese  & 8 \\
CLUENER2020   & 10,748          & 1,343      & 1,345       & 13K       & Chinese  & 10 \\
CNERTA        & 34,102          & 4,440      & 4,445       & 42,987    & Chinese  & 3 \\ \midrule 
Twitter-2015  & 6,176           & 1,546      & 5,078       & 12,784    & English  & 4 \\
Twitter-2017  & 4,290           & 1,432      & 1,459       & 7,181     & English  & 4 \\
\hline
\rowcolor[HTML]{EFEFEF}
{\ourdataset} & 14,678          & 2,059      & 1,842       & 18,579    & Chinese  & 7 \\ \bottomrule
Dataset       & \multicolumn{1}{c}{\# Train doc}    & \multicolumn{1}{c}{\# Dev doc} & \multicolumn{1}{c}{\# Test doc}     & \multicolumn{1}{c}{Size} & Language & \# Class \\ 
\hline\hline
OPP-115        & 75 doc          & /          & 40 doc      & 115 doc   & English  & 12 \\
APP-350        & 188  doc        & 62 doc     & 100 doc     & 350 doc   & English  & 18 \\
\hline
\rowcolor[HTML]{EFEFEF}
{\ourdataset} & 386 doc         & 48 doc     & 49 doc      & 483 doc   & Chinese  & 7 \\ \bottomrule
\end{tabular}%
}
\caption{A comparison between {\ourdataset} and other popular sequence labeling datasets. \# denotes \textit{"number"}. \textit{"doc"} denotes \textit{"documents".}}
\label{tab:data_set_comparison}
\end{table*}

We first compare the size and classes in different datasets.
Table~\ref{tab:data_set_comparison} shows that {\ourdataset} contains abundant semantics, i.e.,  {\ourdataset} has seven annotation classes that are larger than most other datasets (seven out of nine).
%
%
For privacy policy-related datasets, the comparison is conducted with the number of documents as one privacy policy corresponds to one app.
%
OPP-115 annotates at the sentence level, and APP-350 only annotates data controller, data entities, and modifiers.
Since APP-350 specifies data entities into 16 categories, APP-350 exhibits more number of classes than {\ourdataset}.
To summarize, {\ourdataset} is the first and largest Chinese Android privacy policy dataset with abundant semantic labels.

\section{Task and Experiment Setup}
\label{sec:exp}

\subsection{Task description}





{\ourtask} figures out \underline{who} \underline{collects} or \underline{shares} what kind of \underline{data} to \underline{whom}, under which kind of \underline{condition}, and \underline{for what}.
The underlined words correspond to each type of annotations. 
As {\ourtask} concentrates on data access-related sentences, we first locate the sentences based on data collection and sharing words~\cite{ppcheck,yu2016can}.
We summarize the word list based on laws, app market requirements and previous works~\cite{yu2016can,andow2019policylint,ppcheck} (detailed in Appendix~\ref{app:action_list}).
%
%
Given the sentences $C=c_1,c_2,...,c_n$ and its labels $L=l_1,l_2,...,l_n$, where $c_i$ denotes the \textit{i}-th Chinese characters and $l_i$ denotes the $c_i$'s label, the task is to identify  sequence labels.
%
%

\subsection{Summarize models}
\label{sec:model}

This section introduces baseline methods for sequence labeling task on {\ourdataset}.
%
%

\subsubsection{Probabilistic models}
\noindent\textbf{Hidden Markov Model (HMM):} 
%
HMM\footnote{\tiny{\url{https://github.com/luopeixiang/named_entity_recognition}}\label{ner_luopeixiang}}~\cite{freitag2000information} is one of the most classic probabilistic models and is applied as our baselines.  

\noindent \textbf{Condition Random Field (CRF):}
%
%
CRF\footnote{\tiny{\url{http://crfpp.sourceforge.net/}}\label{crfpp_link}}~\cite{lafferty2001conditional} aggregates the advantages of HMM and counters the label bias problems. 
%

\subsubsection{Neural network models}

\noindent\textbf{BiLSTM:}
BiLSTM\textsuperscript{\ref{ner_luopeixiang}}~\cite{graves2005framewise} uses neural network to learn a mapping relation from sentences to labels through the nonlinear transformation in high-dimensional space.
%

\noindent\textbf{BiLSTM-CRF:} BiLSTM-CRF\textsuperscript{\ref{ner_luopeixiang}} uses BiLSTM as a encoder to map the sentences in to a hingh dimension vector and uses CRF as a decoder.
%


\noindent \textbf{BERT-BiLSTM-CRF:}
Since BiLSTM-CRF is still limited to the word vector presentation, BERT-BiLSTM-CRF\footnote{\tiny{\url{https://github.com/macanv/BERT-BiLSTM-CRF-NER}}\label{bertBiLSTMCRF_link}}~\cite{dai2019named} uses BERT as a feature extractor and takes advantage of BiLSTM and CRF for sequence labeling. 

\subsubsection{Lattice enhanced models}

As Chinese words are not naturally separated by space, character-based methods omit the information hidden in word sequences.
Thus, lattice-based methods that integrate lattice information are proposed for Chinese sequence labeling and achieve the promised performance.
%

\noindent \textbf{LatticeLSTM:}
LatticeLSTM\footnote{\tiny{\noindent\url{https://github.com/LeeSureman/Batch_Parallel_LatticeLSTM}}}~\cite{zhang2018chinese}  takes inputs as the character sequence together with all character subsequences that match the words in a predefined lexicon dictionary.

\subsection{Setup and implementation details}

%
%


We evaluate baselines on an Ubuntu 20.04 server with 5 NVIDIA GeForce 3090 (24 GB memory for each), 512 GB memory, and an Intel Xeon 6226R CPU.
Next, we present our implementation details.
For HMM, the number of states, i.e., class number in our dataset with the BIO tag, is set as 22, and the number of observations, i.e., the number of different characters, is set as 1756, which is default value\textsuperscript{\ref{ner_luopeixiang}}.
For CRF, we use the default settings in CRF++\textsuperscript{\ref{crfpp_link}}.
For BiLSTM and BiLSTM-CRF, embedding size is 128,  learning rate is 0.001, and we train models using 30 epochs with a batch size of 64.
For BERT-BiLSTM-CRF\textsuperscript{\ref{bertBiLSTMCRF_link}}, we use the  Chinese bert-base\footnote{\tiny{\url{https://github.com/google-research/bert}}} pretrained model and fine tune it on our training data.
The BiLSTM is set with 128 hidden layer and a learning rate of 1e-5.
BERT-BiLSTM-CRF model is trained on our dataset with default settings\textsuperscript{\ref{bertBiLSTMCRF_link}} where the batch size is 64, learning rate is $1e^(-5)$, dropout rate is 0.5, gradient clip is 0.5, and early stop strategy is "\textit{stop if no decrease}".
For Lattice-LSTM, we use the same lattice provided in ~\cite{zhang2018chinese}.







\section{Evaluation}
\label{sec:evaluation}

In this section, we evaluate baseline methods on all 18,579 sentences that are divided into training, development, and testing sets as detailed in Table~\ref{tab:data_set_comparison}.
Following previous research~\cite{wilson2016creation, sui2021large}, we apply precision (P), recall (R), and F1-score (F1) to evaluate baselines.



%
Table~\ref{tab:main_results} shows the overall performance of families of baselines on {\ourdataset}.
Table~\ref{tab:main_results} shows that BiLSTM-CRF achieves the most promising performance, which may benefit from the enhanced presentation ability of bidirectional LSTM and CRF for capturing the context information.
%
%
Lattice-LSTM performs a strong representation of capturing lattice information, while some clauses in our labels may mislead the model learning the patterns.

\begin{table}[]
\centering
\resizebox{0.89\linewidth}{!}{%
\begin{tabular}{@{}cccc@{}}
\toprule
\multicolumn{1}{c}{} & P & R & F1 \\ \midrule
HMM & 77.47\% & 66.11\% & 69.63\% \\
CRF & 85.52\% & 86.28\% & 85.63\% \\
BiLSTM & 85.13\% & 85.99\% & 85.05\% \\
BiLSTM-CRF & 86.94\% & 86.90\% & 86.84\% \\
BERT-BiLSTM-CRF & 46.22\% & 57.35\% & 51.18\% \\
Lattice-LSTM & 78.63\% & 80.75\% & 79.67\% \\
\bottomrule
\end{tabular}%
}
\caption{Overall performance of baseline methods on our dataset.}
\label{tab:main_results}
\end{table}


We analyze the identification performance of each component to investigate the challenges and limitations of  {\ourdataset}.
Table~\ref{tab:detail_results} demonstrates the detailed performance of baselines, i.e., CRF-based models, BERT-based models, and Lattice-based models.
Besides, we also compare the performance with \textit{manual agreements} to demonstrate task difficulties. 
%
Table~\ref{tab:detail_results} demonstrates that BiLSTM-CRF and Lattice-LSTM achieve over 90\% performance on \textit{Receiver} because the \textit{Receiver} possesses few overlaps with other labels and is in the format of words.
\textit{Collect} and \textit{share} only achieve around 60\% precision and F1-score because the two types of entities perform some overlapping, as is shown in Fig.\ref{fig:ann_demo} and Fig.\ref{fig:app_overlap}.
Table~\ref{tab:detail_results} shows that BiLSTM-CRF achieves better precision on \textit{Condition}  
than Lattice-LSTM, which may be caused by the fact that \textit{Condition} and \textit{Purpose} are mainly in the format of attributive clauses rather than words.

\begin{table*}[]
\centering
\resizebox{0.98\textwidth}{!}{%
\begin{tabular}{@{}cccc|ccc|ccc|ccc@{}}
\toprule
& \multicolumn{3}{c|}{BiLSTM-CRF} & \multicolumn{3}{c|}{BERT-BiLSTM-CRF} & \multicolumn{3}{c|}{LaticeLSTM} & \multicolumn{3}{c}{Manual Agreements} \\ \midrule
& P        & R        & F1       & P          & R          & F1        & P        & R        & F1       & P           & R          & F1         \\ \midrule
\multicolumn{1}{c|}{Collect}    & 51.80\%  & 57.50\%  & 54.47\%  & 50.59\%    & 68.89\%    & 58.34\%   & 69.23\%  & 67.05\%  & 65.10\%  & 96.30\%     & 92.07\%    & 94.14\%    \\
\multicolumn{1}{c|}{Condition}  & 81.75\%  & 72.76\%  & 77.00\%  & 31.59\%    & 46.46\%    & 37.61\%   & 72.76\%  & 77.00\%  & 81.75\%  & 93.53\%     & 84.50\%    & 88.79\%    \\
\multicolumn{1}{c|}{Data}       & 77.85\%  & 58.60\%  & 66.44\%  & 51.11\%    & 67.19\%    & 58.06\%   & 58.60\%  & 66.44\%  & 77.85\%  & 96.20\%     & 91.79\%    & 93.94\%    \\
\multicolumn{1}{c|}{Controller} & 64.10\%  & 61.08\%  & 62.50\%  & 56.53\%    & 63.80\%    & 59.94\%   & 61.08\%  & 62.50\%  & 64.10\%  & 96.96\%     & 90.18\%    & 93.45\%    \\
\multicolumn{1}{c|}{Purpose}    & 70.88\%  & 54.61\%  & 60.64\%  & 40.45\%    & 48.46\%    & 44.09\%   & 54.61\%  & 60.64\%  & 70.88\%  & 95.64\%     & 92.61\%    & 94.10\%    \\
\multicolumn{1}{c|}{Share}      & 68.31\%  & 51.83\%  & 58.88\%  & 59.08\%    & 45.61\%    & 51.48\%   & 51.83\%  & 58.88\%  & 68.31\%  & 96.10\%     & 94.71\%    & 95.40\%    \\
\multicolumn{1}{c|}{Receiver}   & 91.70\%  & 92.68\%  & 92.19\%  & 22.96\%    & 27.84\%    & 25.17\%   & 92.68\%  & 92.19\%  & 91.70\%  & 97.33\%     & 85.00\%    & 90.75\%    \\
\multicolumn{1}{c|}{O}          & 91.70\%  & 92.68\%  & 92.19\%  & 46.22\%    & 57.35\%    & 51.18\%   & 92.57\%  & 92.79\%  & 92.35\%  & /           & /          & /          \\
\multicolumn{1}{c|}{Average}    & 86.94\%  & 86.90\%  & 86.84\%  & 37.54\%    & 49.42\%    & 42.66\%   & 72.27\%  & 72.27\%  & 72.27\%  & 96.01\%     & 90.12\%    & 92.94\%    \\ \bottomrule
\end{tabular}%

}
\caption{Evaluation performance of three types of methods on our dataset. "O" denotes \textit{others}.}
\label{tab:detail_results}
\end{table*}

\begin{figure}[]
    \centering
    \includegraphics[width=0.95\linewidth]{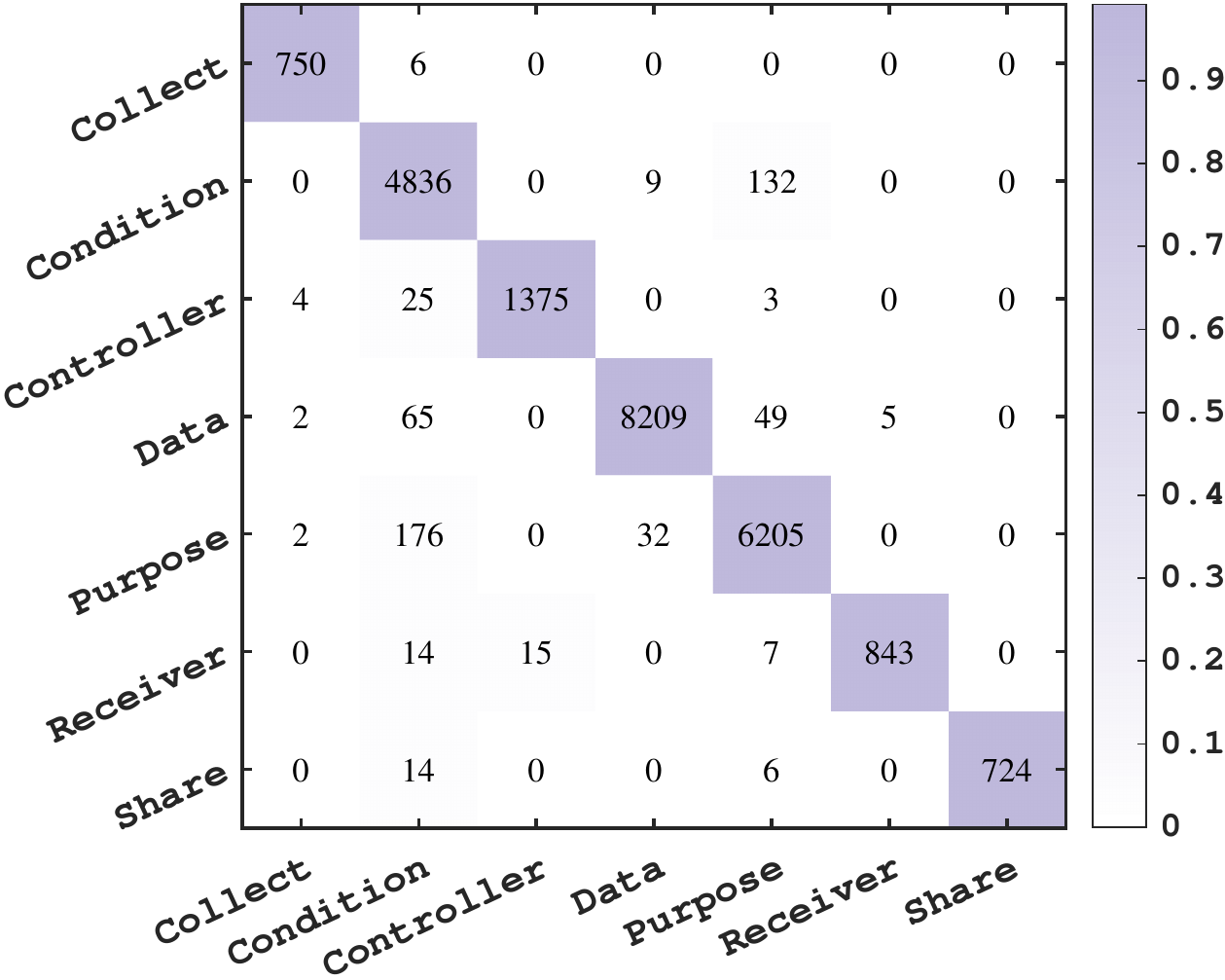}
    \caption{Confusion matrix of BiLSTM-CRF results on {\ourdataset}.}
    \label{fig:confusion_matrix_results}
\end{figure}

Next, we analyze the confusion matrix of BiLSTM-CRF results that performs the best on {\ourdataset}.
%
In Fig.\ref{fig:confusion_matrix_results}, the depth of background color denotes the proportion of classification, the darker the color the higher the proportion, and the digit denotes the number of classification results.
Fig.\ref{fig:confusion_matrix_results} indicates that most of the misclassification samples are related to \textit{Condition}.

\begin{figure}[]
    \centering
    \subfigure[Missing condition.]{
        \centering
        \includegraphics[width=0.95\linewidth]{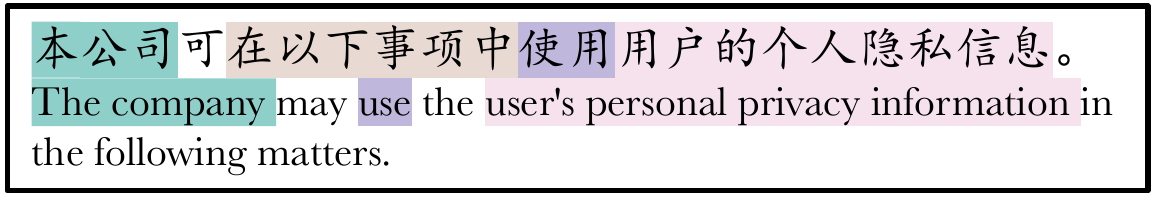}
        \label{fig:miss_condition}
    }

    \subfigure[Error prediction when controller is user.]{
        \centering
        \includegraphics[width=0.95\linewidth]{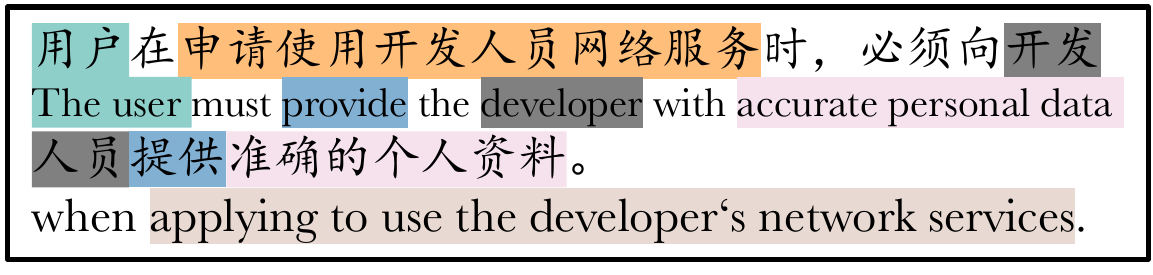}
        \label{fig:user_error}
    }
    \caption{The visualization of divergence between ground truth and prediction.}
\end{figure}

To have a deep understanding of divergences between ground truth and predictions, we inspect the misclassifications. 
We find that the algorithm may fail to identify \textit{Condition}s, which are in the adverbial clause as shown in Fig.\ref{fig:miss_condition} where the highlighting for Chinese is ground truth and highlighting for English is prediction results.
Besides, when the data controller is the user, as is shown in Fig.\ref{fig:user_error}, the algorithms fail to distinguish \textit{Purpose} and \textit{Condition}.
More illustrations in Appendix~\ref{app:demo} also reveal that models need to be well designed to learn deep semantic information, such as distinguishing overlapping among components, and distinguishing \textit{Purpose} in modifiers.

\section{Case Study}
\label{sec:case}

In this section, we will present cases of potential applications of {\ourdataset}, such as whether privacy policies comply with regulatory requirements and whether privacy policies is consistent with the apps' functionalities.

\begin{figure}[]
    \centering
    \includegraphics[width=0.93\linewidth]{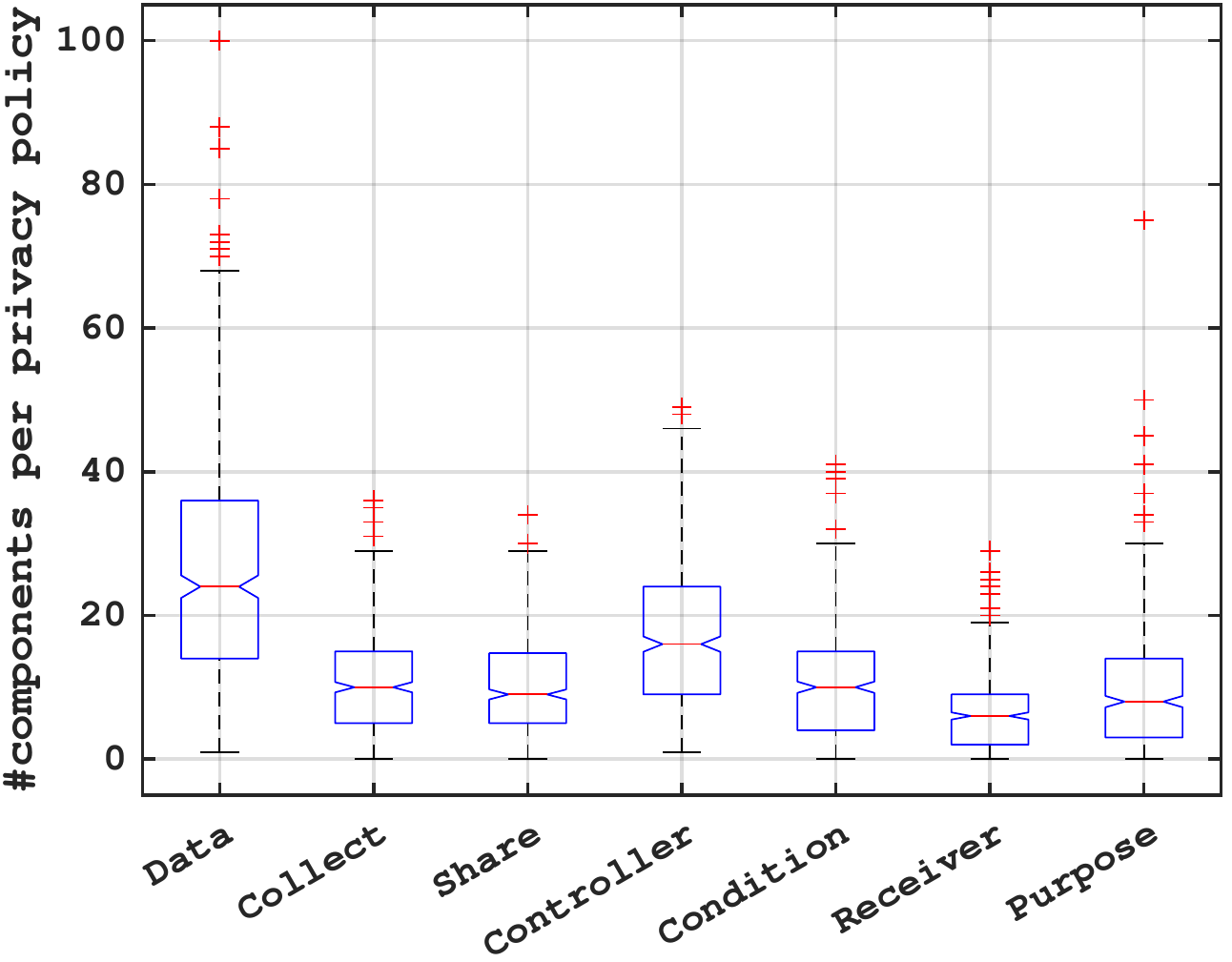}
    \caption{Components distribution of {\ourdataset}.}
    \label{fig:components_distribution}
\end{figure}

\noindent \textbf{Regulation compliance identification}.
Chinese privacy-related laws~\cite{piss,zhSDKLaw,clprc} ask developers to clearly claim purpose
conditions for processing user privacy data.
%
We first investigate the distribution of annotations in {\ourdataset}.
Fig.\ref{fig:components_distribution} sketches the box plot of the frequency of components in each privacy policy. 
Fig.\ref{fig:components_distribution} indicates that some privacy policies claim data processing without clarifying the purpose and condition, i.e., the minimum frequency of \textit{Data} is positive while that of \textit{Purpose} is zero.
We manually inspect privacy policies. 
%
We find that the privacy policies, whose package name is \textit{com.yitong.weather}, 
claim the app collects users' data while omitting to give the purposes or conditions of data access, which violates regulation requirements.
Thus, {\ourdataset} can facilitate the research in the area of privacy compliance identification~\cite{andow2019policylint, barth2022understanding}.

\noindent \textbf{App behavior consistency identification.}
To improve the security of the Android community, researchers design systems~\cite{ppcheck,ppchecker} to identify the consistency between privacy policies and app behaviors to prevent apps from abusing user data or conducting malicious behavior.
One popular method to check the app's behavior is dynamic analysis~\cite{yan2012droidscope}, i.e., running the app on the device and checking the \textit{log} information. 
%
To investigate the application of {\ourdataset} in security community, we first identify the privacy policies without \textit{purpose} or \textit{condition} components.
Then, we install the app on one smartphone, manually interact with the app and try our best to trigger all possible functions in the app by clicking every visible buttons.
We use \textit{logcat} to capture the app's running information.
We find that the app (id: \textit{com.chengmi.signin}) requests device storage to use the app's functionalities while no condition-related statements are claimed in its privacy policy.
With more intelligent automatic software engineering techniques, {\ourdataset} can facilitate the research in this area, and more vulnerabilities in the consistency between app behavior and privacy policy could be investigated.



\section{Discussion}
\label{sec:dis}

In this section, we first discuss difficulties in {\ourdataset}.
Then, we propose potential research topics on {\ourdataset}.
Finally, we discuss limitations of {\ourdataset}.
Besides, we also discuss ethical concerns in Appendix~\ref{app:ethical}.

\subsection{Dataset difficulties}

%
Based on evaluation results in \S\ref{sec:evaluation} and related work, we raise the following difficulties: 
%
1) How to distinguish overlaps between components?
2) How to effectively deal with length variation of 
components?
3) Difficulties in semantic analysis.

\begin{CJK}{UTF8}{gbsn}
Different from traditional sequence labeling tasks, components in our data set may contain other components.
One scenario is the Purpose or Conditions maybe used to decorate the data, for example, "\textit{We will collect your login information} (我们会收集您的登录信息)" where the \textit{login} may be understood as the purpose of \textit{information}. 
Since traditional sequence labeling methods predict one character with one label, it is hard to distinguish components overlaps.
One possible solution is using multi-model algorithms~\cite{sui2021large} that demonstrate effectiveness for distinguishing boundaries between entities. 
Similar to traditional news or social media datasets that use voice or images as additional information, integrating apps' analysis results 
help distinguish different components. 
%
\end{CJK}

Second, existing sequence labeling tasks mainly concentrate on entity recognition, while practical  applications may require labeling clauses for further analysis.
Table~\ref{tab:data_statistic} shows that average length of components in {\ourdataset} varies from 2.03 to 19.24.
{\ourtask} not only require identifying words but also ask the models to identify the role of clauses.

The semantic analysis of privacy policies is still a difficulty.
%
Laws require apps to clearly clarify how apps collect and share user data.
%
Privacy policies can claim that apps will \textit{share} data with third parties or third parties will \textit{collect} user data.
In this way, it becomes essential to understand the context to distinguish the controller and action type.
It could be a solution to use multi-model algorithms integrating program analysis to improve the performance; however, identifying the third party and app itself remains a challenge in program analysis.


\subsection{Further directions}

The {\ourdataset} enables research in directions of interest to natural language processing, privacy protection, and cyber security~\cite{zhu2022adversarial,zhu2022binarizedattack}.
We propose some potential research interests for further work below.

\noindent\textbf{Emotional analysis in privacy policies.}
Existing research~\cite{andow2019policylint} figures out privacy policies may conflict among contexts.
For example, the privacy policy may claim NOT to collect user data in one sentence while claiming to access user data in other sections.
Existing methods~\cite{andow2019policylint, ppcheck,ppchecker} use negative words to identify whether conflicts exist in the privacy policy and ignore complications like a double negative.
In Chinese privacy policy, negative representations are more complicated~\cite{liu2012sentiment}.
Thus, emotional analysis can help analysts better understand the semantics of privacy policies..

\noindent\textbf{Privacy compliance detection.}
{\ourdataset} provides detailed labels for data usage, including the purpose and conditions.
It is necessary to investigate the detailed requirements of laws and further identify whether existing privacy policies violation.

\noindent\textbf{Cyber security investigation.}
Privacy policies ought to reflect the functionalities of apps.
Some apps may conceal the malicious behavior in their functionalities and do not claim the behavior in privacy policies.
{\ourdataset} can help identify the consistency between apps' functionalities and behavior by combing natural language process algorithms and code analysis.

\begin{CJK}{UTF8}{gbsn}

\end{CJK}

\subsection{Limitations}

{\ourdataset} provides detailed annotations for data access statements in privacy policies.
However, 
analyzing privacy policies using {\ourdataset} depends on the performance of locating data access-related sentences.
We use data collection and sharing words to locate the sentences.
However, some Purpose and Condition claims maybe given as an enumeration format, such as "\textit{we will not share your personal data under the following conditions}".
{\ourdataset} is limited when capturing  information in enumeration format.

Privacy policies possess timeliness.
App developers should provide a privacy policies when publishing the apps.
When the apps' functionality updates, the privacy policies ought to be updated accordingly.
The data set is limited to the timestamp we collected.
When combining our dataset with program analysis, this factor should be considered.


\section{Related Work}
\label{sec:relatedw}

Prior privacy policy datasets are all English and omit other languages.
OPP-115~\cite{wilson2016creation} collects 115 English websites' privacy policies and makes annotations at the sentence level.
OPP-115 designs labels based on previous works~\cite{mcdonald2008cost,staff2011protecting}. 
%
APP-350~\cite{zimmeck2019maps} gathers Android apps' privacy policies written in English.
APP-350 only conducts limited annotations, including two types of data controllers, namely first party and third party,  thirteen types of specific data, and two types of modifiers, i.e., do and do not.

Existing Chinese sequence labeling datasets are generally gathered from News~\cite{zhang2006word,peopodaily,sui2021large} and social media~\cite{peng2016improving, weischedel2011ontonotes, zhang2018chinese}.
The datasets include abundant corpus, but their annotations are limited to location, person name, and organization.
Even though CLUENER2020~\cite{xu2020cluener2020} expands the labels, such as game, gvoerment, book,  the datasets are still hard to be applied in specific downstream tasks.
%
CNERTA~\cite{sui2021large} includes another media data, i.e., voice data, to improves the sequence labeling performance.







\section{Conclusion}
\label{sec:conclusion}

This paper introduces the first Chinese Android application privacy policy dataset, {\ourdataset}.
{\ourdataset} contains fine-grained annotations based on requirements of privacy-related laws and regulations.
The dataset can help promote natural language processing research on practical downstream tasks.
We also conduct experimental evaluations of popular baselines on our dataset and propose potential research directions based on the result analysis.
We also conduct case studies to investigate potential applications of our dataset and potential application of our dataset to help software engineering and cyber security protect user privacy.
In the future, we hope that we can build new models for {\ourdataset} to counter the challenges. 
%
















\section*{Acknowledgements}
We thank the anonymous reviewers for their insightful comments.
This work was partially supported by Hong Kong RGC Project (PolyU15224121), NSFC Young Scientists Fund (62006203), and HKPolyU collaborative research project (ZGBE).

\bibliographystyle{acl_natbib}
\bibliography{custom}

\section{Appendix}
\label{sec:appendix}

\subsection{Data access word list}
\label{app:action_list}

Table~\ref{tab:wordlist} gives data sharing and collection word list, that is summarized from laws~\cite{zhSDKLaw,gdpr,piss}, app market requirements~\cite{googleplay_requirements,appgallery_requirements}, and previous works~\cite{yu2016can,andow2019policylint,ppcheck}.
With those words, researchers can locate data access-related sentences and conduct further analysis to get interested entities, such as data controller, data entity, collection, sharing, condition, purpose and data receiver.


\begin{table}[]
\centering
\begin{CJK}{UTF8}{gbsn}
\resizebox{0.85\linewidth}{!}{%
\begin{tabular}{p{0.23\columnwidth}p{0.67\columnwidth}}
\toprule
Sharing & 收集 (collect), 获取 (obtain), 接受 (get), 接收 (receive), 保存 (save), 使用 (use), 采集 (gather), 记录 (record), 存储 (store), 储存 (store) \\
\hline
Collection & 披露 (reveal),	分享 (share), 共享 (share), 交换 (exchange), 报告 (report), 公布 (public), 发送 (send), 交换 (exchange), 转移(transfer), 迁移 (migrate), 转让 (make over),	公开 (public), 透露 (disclose), 提供 (provide) \\ \bottomrule
\end{tabular}%
}
\end{CJK}
\caption{Data access word list}
\label{tab:wordlist}
\end{table}

\subsection{Ethical Consideration}
\label{app:ethical}

{\ourdataset} is a dataset constructed by gathering publicly available privacy policy websites without posing any ethical problems.
First, privacy policies are publicly accessible in multi ways.
According to application markets' requirements, developers or companies are asked to provide those privacy policy websites once they publish their apps.
Privacy policies also ought to be given when the users use apps for the first time according to law requirements~\cite{piss}.
Second, we do not collect any privacy information.
Besides, the {\ourdataset} is proposed to prompt research for protecting user privacy.

For the annotations, we hired part-time research assistants from our university to label the dataset.
They are compensated with 9 USD/hour and at most 17.5 hours per week.




\begin{figure}[t]
    \centering
    \includegraphics[width=1\linewidth]{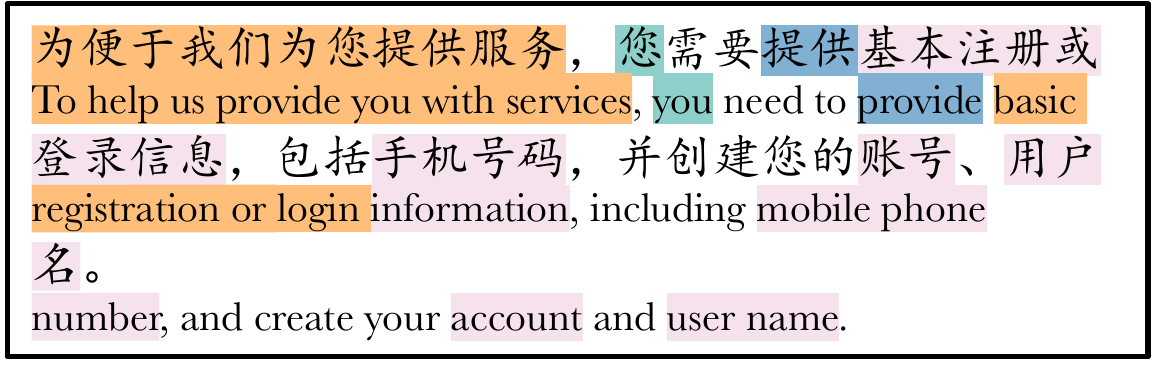}
    \caption{Overlapping between components. Differences between ground truth and prediction.}
    \label{fig:app_overlap}
\end{figure}

\subsection{Prediction results analysis}
\label{app:demo}

In this section, we show the prediction results of the algorithm and some common problems.
These problems could be the limitations of existing models and also be challenges for designing algorithms for our data scenario.

\begin{figure}[t]
    \centering
    \includegraphics[width=0.95\linewidth]{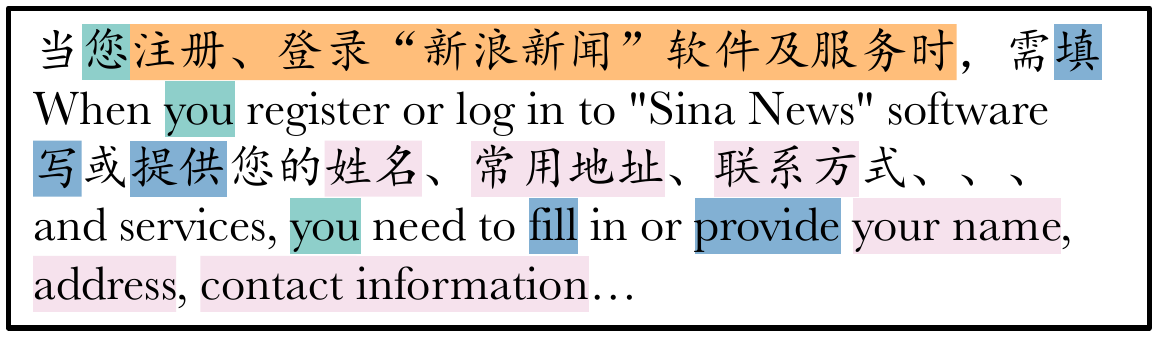}
    \caption{The visualization of divergence between ground truth and prediction for missing Purpose.}
    \vspace{-0.5em}
    \label{fig:user_purpose}
    \vspace{-1em}
\end{figure}

\begin{CJK}{UTF8}{gbsn}
Fig.~\ref{fig:app_overlap} illustrates the scenario where there exist overlapping between components, i.e., the "\textit{basic registration or login information} (基本注册或登录信息)".
Exactly, "basic registration or login information" should be one data as is highlighted in Chinese version, i.e., the ground truth.
However, the algorithm will prediction "\textit{basic registration or login} (基本注册或登录)" as Purpose and "\textit{information}(信息)" as Data as are highlighted in English version.
\end{CJK}
The meaning of color for different categories can be refer to Figure~\ref{fig:ann_demo}.
Fig.~\ref{fig:user_purpose} shows the pre-trained algorithm may misclassify \textit{Purpose} as \textit{Condition} when data controller is the user.
%


\end{document}